# CREDIBLE DISCOURSE ON CLIMATE POLICY: Beyond Dueling Certitudes

Charles F. Manski

Department of Economics and Institute for Policy Research
Northwestern University

September 2026

Abstract

Whatever the policy question under consideration, reasoned and realistic evaluation requires credible policy analysis under uncertainty. This holds particularly to the study of climate policy in the United States, where science and ideology have become increasingly entangled. Welfare economics provides a transparent formal framework to describe and evaluate the subtle tradeoffs that must be addressed when comparing alternative climate policies. The largely qualitative discussions of climate policy in public-facing reports suffer from failure to use welfare economics to assess tradeoffs. I use 2025 reports by the National Academies and by the US Department of Energy as case studies.

## 1. Introduction

Present discourse on climate policy in the United States is unhealthy, with non-constructive debate between mainstream and contrarian perspectives. Dispute about the causes and consequences of climate change has become akin to a religious war, with productive communication seemingly impossible.

Mainstream climate science, exemplified by the Sixth Assessment Report of the Intergovernmental Panel on Climate Change (IPCC), has concluded that climate change is a serious planetary problem that warrants concerted societal efforts to slow and eventually stop global warming (IPCC, 2023). The prevailing thrust of climate science has been summarized in a March 2026 Statement released by the American Meteorological Society, which asserts that[1]

- "Climate is changing, and the rate and magnitude of change are unusual in relevant human experience."
- "People are the primary cause of modern global climate change, including through burning coal, oil, and natural gas."
- "Climate change is harmful overall, and the threats to people and all life are increasing."
- "Those who study the scientific evidence overwhelmingly agree."

A minority scientific position, sometimes described as contrarian (e.g., Voosen, 2025), asserts that global warming is not a problem requiring development of assertive public policies to lessen greenhouse gas (GHG) emissions. This view gained public visibility in July 2025 when the U.S. Department of Energy (DOE) released a report authored by five climate scientists with reputations as contrarians (Climate Working Group, 2025). The report discussed mainstream climate research and agreed that global warming is occurring, but it disputed the causes and consequences. The Executive Summary states: "models and experience suggest that $CO_2$-induced warming might be less damaging economically than commonly believed, and excessively aggressive mitigation policies could prove more detrimental than beneficial . . . U.S. policy actions are expected to have undetectably small direct impacts on the global climate."

[1] See https://www.ametsoc.org/ams/about-ams/ams-statements/statements-of-the-ams-in-force/statement-on-public-availability-of-scientific-information-and-scientific-evidence-on-climate-change/, accessed March 24, 2026.

In September 2025, the National Academies of Sciences, Engineering, and Medicine (NASEM) released a Consensus Study Report authored by sixteen climate, environmental, and other researchers. Most NASEM reports are written in response to a request from an agency of the federal government. However, this "fast-track" study was self-commissioned in response to a notice of proposed rulemaking issued by the U.S. Environmental Protection Agency (EPA) indicating its intention to rescind its 2009 "Endangerment Finding," which provided the legal basis for federal climate regulations under the Clean Air Act from 2009 until early 2026.[2] Expressing the perspective of mainstream climate science, the NASEM report reached this Overarching Conclusion (p. 1): "EPA's 2009 finding that the human-caused emissions of greenhouse gases threaten human health and welfare was accurate, has stood the test of time, and is now reinforced by even stronger evidence." The NASEM report cited no contrarian research, including none by any authors of the DOE report. It only mentioned the DOE report briefly (page 7) and did not discuss its content.

### 1.1. Incredible Certitude in Climate Discourse

Formation of a reasonable climate policy would be challenging even in an ideal world where researchers and the public communicate in a coherent, clear, and respectful manner. Alternative policies may have complex societal impacts, which may be highly uncertain. Research may never yield complete knowledge of policy impacts. Disagreements regarding the values that society should pursue may persist.

These deep issues notwithstanding, the present state of climate discourse is unnecessarily dysfunctional. Careful welfare economic study of policy choice under uncertainty can reduce the misunderstandings and inconsistencies that often afflict policy debates. Manski (2013, 2024, 2026) use welfare economics to analyze a spectrum of complex policy issues. Manski, Sanstad, and DeCanio (2021) and Decanio, Manski, and Sanstad (2022) use it to study climate policy under uncertainty.

[2] See https://www.govinfo.gov/content/pkg/FR-2026-02-18/pdf/2026-03157.pdf, accessed March 27, 2026.

Unfortunately, policy analysis rarely addresses fundamental uncertainties. Researchers commonly express certitude about the consequences of alternative policies, making firm predictions resting on unsubstantiated assumptions. They have thus performed policy analysis with *incredible certitude* (Manski 2011, 2019, 2020).

Some of the practices contributing to incredible certitude are

- conventional certitude: A prediction generally accepted as true but not necessarily true.
- dueling certitudes: Contradictory predictions made with alternative assumptions.
- conflating science and advocacy: Specifying assumptions to generate a predetermined conclusion.

American discourse on climate change exhibits all these aspects of incredible certitude. To illustrate,

- Conventional certitude is evident in the statement by the American Meteorological Society quoted above, expressing the mainstream position that "Those who study the scientific evidence overwhelmingly agree."
- Dueling certitudes is evident in the unreconcilable differences between mainstream and contrarian climate research. The 2025 DOE report firmly rejects policy prescriptions relying on mainstream climate research. The 2025 NASEM report ignores the arguments made by contrarian researchers.
- Conflating science and advocacy is indicated in the choice made by the DOE to commission a report written solely by five contrarian authors and in the choice made by NASEM not to include contrarian researchers in its committee.

Recognizing uncertainty will not eliminate scientific disagreements and differences of perspective on societal values, but it should enable productive communication across researchers and amongst the public. Section 2 observes that welfare economics provides a coherent framework to describe and evaluate the subtle tradeoffs that must be addressed when comparing alternative climate policies. Section 3 uses the 2025 NASEM and DOE reports as case studies to illustrate how discourse on climate policy in highly visible public-facing reports suffers from failure to use welfare economics to assess tradeoffs.

## 2. Welfare Economic Analysis of Climate Policy

Climate change is a complex dynamic physical phenomenon with a multitude of potential societal implications. Hence, credible welfare economic analysis of climate policy is challenging. Nevertheless, the basic principles are easy to understand.

Economists have long used the construct of a social welfare function (SWF) to compare policies. Research in public economics seeks to characterize the social welfare achieved by alternative policies, aiming to find one that maximizes an SWF. Economists studying policy choice in democracies strive to specify SWFs that in some manner express the values of population members rather than the preferences of dictators.[3] The public at large may not be familiar with the formal structure of welfare economics, but basic ideas are familiar through the widespread use of the term *benefit-cost analysis*.

Welfare economic analysis becomes operational when several questions are answered: What constitutes social welfare? What are the feasible actions? What is known about the welfare consequences of alternative choices?

If the available knowledge suffices, economists seek to determine an optimal policy. If uncertainty prevents optimization, they seek to determine a policy that satisfies some criterion for reasonable decision making under uncertainty. Most prevalent has been maximization of expected social welfare. One places a subjective probability distribution on unknown quantities and chooses an action that maximizes the expected value of welfare with respect to this distribution. When it is not credible to express uncertainty through a subjective probability distribution, researchers apply criteria that, in some sense, work uniformly well over all of feasible values of the unknown quantities. Two prominent interpretations of this broad idea are the maximin and minimax-regret criteria. See Manski (2024) for extended discussion and applications.

Climate research performed by earth scientists is an input into welfare economic analysis of climate policy. Welfare economic analysis is not within the domain of expertise of earth scientists. Such analysis has been the mission of climate economists who use i*ntegrated assessment* (IA) models to evaluate climate policies. IA models are long-run (century-scale or more) descriptions of the global economy, including the energy system and its role in economic production. These models represent the climate and the links between the climatic effects of GHG emissions and the economy. A prominent example is Nordhaus' Dynamic Integrated Climate Economy (DICE) model (Nordhaus, 2019).

---

[3] For example, the U.S. Constitution begins with the words "We the People" and refers to the *general Welfare*.

Policy comparisons are performed by considering a planner who seeks to make an optimal trade-off between the costs of GHG abatement and the economic damage from climate change. The planning problem is commonly formalized as an optimal-control problem with these components:

(1) equations coupling GHG emissions and abatement to the accumulation of GHGs in the atmosphere and resulting temperature increases.
(2) a damage function quantifying the societal effects of climate change as a function of temperature increases.
(3) an abatement cost function that expresses the cost of actions to reduce GHG emissions relative to a stipulated baseline emissions trajectory.

The first component is based on research in earth science. The second and third components quantify the economic aspects of a specified SWF.

IA models make plain the fundamental difficulty in choosing between moderate and aggressive policies to reduce GHG emissions. Moderate (aggressive) polices have lower (higher) abatement costs. Moderation yields greater (less) global warming and hence greater (smaller) potential damage from climate change. Appropriately quantifying these opposing tendencies is essential to optimal policy choice.

Studying climate policy as a problem of optimal control presumes that a planner knows enough about the climate and economic systems to make optimization feasible. Yet uncertainties abound. Physical and economic uncertainties have been handled in different ways. I describe below.

## 2.1. Multi-Model Ensembles of Climate Models

The climate is a complex system comprising many physical processes occurring at a range of spatial and temporal scales. Climate models aim to represent these processes in a tractable manner. These models contain equations describing large-scale atmospheric dynamics. Uncertainty arises in part because implementation of the equations in models requires numerous practical choices involving discretization and solution methods. Moreover, some components of the system, such as cloud formation and heat transfer between land surfaces and the atmosphere, are not yet fully understood and must be approximated.

Multiple climate models have been developed, each reflecting different but credible choices in model design and implementation. These models yield different forecasts of the global climate. Neither a consensus climate model nor definitive quantitative forecasts can be specified with current knowledge (Pindyck, 2022). The range of forecasts produced by different climate models indicates current uncertainty about the climate system.

Seeking to eliminate uncertainty, climate scientists have performed *multi-model ensemble* (MME) analysis (Taylor et al., 2012). MME analysis usually combines model outputs into a single forecast of future climate variables. Modelers have perceived policymakers as requiring single forecasts, as functions of particular GHG emissions scenarios, for use in decision-making (Parker 2006). However, climate researchers have recognized persistent problems in combining forecasts (Tebaldi and Knutti, 2007; Sanderson, 2018).

A common technique is to take the simple average across model forecasts of policy-relevant variables such as increases in global mean temperature due to GHG emissions. Computation of a simple average of predictions, sometimes called *model democracy*, gives equal weight to each model, an idea lacking a compelling foundation (Knutti, 2010). Researchers may instead compute weighted average forecasts when they believe that models can be ranked with respect to relative accuracy.

Climate scientists recognize that combining climate model ensemble outputs into a single projected trajectory of the future global climate remains a challenging and unresolved problem. As summarized in an IPCC physical sciences report, "…despite some progress, no universal, robust method for weighting a multi-model forecast ensemble is available…" (Lee et al., 2021). This poses a quandary for policymakers who rely on climate model forecasts to formulate strategies for GHG emissions abatement and other approaches to address climate change.

Manski, Sanstad, and DeCanio (2021) argued that MME analysis be abandoned. To use climate forecasts in IA welfare economic policy assessment, it was proposed to frame climate model uncertainty as a problem of *partial identification*, or *deep uncertainty*. This terminology refers to situations in which the underlying mechanisms, dynamics, or laws governing a system are not completely known and cannot be credibly modeled definitively even in the absence of data limitations in a statistical sense. The *minimax regret* (MMR) decision criterion was suggested to account for deep climate uncertainty, without weighting climate model forecasts. A theoretical framework for cost-benefit analysis of climate policy based on MMR was developed and applied computationally.

## 2.2. Economic Uncertainties

Climate economists whose research specifies damage functions and abatement cost functions for IA of climate policy have commonly proceeded with incredible certitude. In general, economists have not performed MME analyses. They have instead reported disparate findings, stemming from their separate studies.

DeCanio, Manski, and Sanstad (2022) generalized the MMR analysis of Manski, Sanstad, and DeCanio (2021) to encompass physical-science uncertainty regarding the correct climate model and one among multiple economic uncertainties. Among the many economic aspects of IA models that have lacked consensus, perhaps the most contentious has been how a planner should assess the costs and benefits of policies across generations. To confront this issue, the paper studied choice of climate policy that minimizes maximum regret with uncertainty regarding both the correct climate model and the appropriate intergenerational assessment of policy consequences.

Economists have long framed intergenerational policy assessment using a time discount rate. They have evaluated climate policies by the present discounted value of the sum of abatement costs and the corresponding damages. There has been considerable and unresolved debate about what discount rate to use; see, for example, Arrow et al. (2014) and Heal and Milner (2014). The choice is highly consequential. Low discount rates favor policies that reduce GHG emissions aggressively and rapidly (Emmerling et al., 2019). High rates favor policies that act more modestly and slowly. To express uncertainty and possible normative disagreement, it was supposed that the appropriate discount rate lies within an interval that covers the spectrum of rates that have been used in the literature.

The IA model studied in Decanio, Manski, and Sanstad (2022) was relatively simple and tractable, in part because it did not consider all potentially important sources of uncertainty. The appropriate baseline emissions path is highly uncertain. The shapes and parameters of the abatement cost and climate damage functions are also uncertain. A further limitation of this work is that it followed the convention in IA modeling of assuming that present discounted gross world product expresses social welfare. This measure of welfare ignores the cross-sectional distribution of personal welfare in the population.

## 3. The Inadequacy of Qualitative Policy Comparison: The NASEM and DOE Reports as Case Studies

There is a fundamental difficulty in choosing between moderate and aggressive policies to reduce GHG emissions: Moderate policies yield lower abatement costs but greater potential damage from global warming. Aggressive policies have contrasting relative costs and benefits. IA models quantify these opposing tendencies, enabling coherent policy choice.

A continuing problem in public discourse on climate policy is the prevalence of qualitative comparison of policies, without quantification of costs and benefits. This practice has contributed to dueling certitudes between the mainstream and contrarian positions. Mainstream reports emphasize the potential damage from warming, downplaying or ignoring the cost of abatement. Contrarian reports do the opposite, emphasizing the cost of abatement while downplaying or dismissing the potential damage from warming. Although mainstream and contrarian reports differ sharply in their policy recommendation, they share the flaw of not dealing with uncertainty appropriately.

The recent NASEM and DOE reports exemplify the shortcomings of qualitative policy comparison, each in its own way. Viewing them as instructive case studies, I discuss them here. I begin with the NASEM report. The NASEM study was performed after public release of the DOE report, but it mentioned the DOE report only briefly and did not discuss its content. Indeed, the NASEM report discussed no contrarian research.

### 3.1. The NASEM Report

The Committee that wrote NASEM (2025) was asked to respond to this Statement of Task (p. 5):

> "This fast-track study will review evidence for whether anthropogenic emissions of greenhouse gases to the atmosphere are reasonably anticipated to endanger public health and welfare in the United States. The study will focus on updates since the Environmental Protection Agency finalized the Endangerment Finding in 2009, examine how current understanding compares to the 2009 Endangerment Finding, and provide explanation for any changes. The study will develop conclusions that describe supporting evidence, the level of confidence, and areas of disagreement or unknowns."

Observe that the Statement of Task did not ask the Committee to assess the magnitude of the severity with which climate change endangers public health and welfare. It was only asked to draw a yes/no conclusion: Does climate change endanger the United States or not? Restriction of the task to drawing a binary conclusion may reflect the fact that the 2009 EPA Endangerment Finding was itself similarly binary. The Endangerment Finding empowered EPA to regulate GHG emissions, but it did not provide guidance on how the costs and benefits of alternative regulations should be evaluated.

I earlier cited the report's main finding (p. 1): "EPA's 2009 finding that the human-caused emissions of greenhouse gases threaten human health and welfare was accurate, has stood the test of time, and is now reinforced by even stronger evidence." This statement was supported by these five conclusions (p.1-2):

> "(1) Emissions of greenhouse gases from human activities are increasing the concentration of these gases in the atmosphere."
>
> "(2) Improved observations confirm unequivocally that greenhouse gas emissions are warming Earth's surface and changing Earth's climate."
>
> "(3) Human-caused emissions of greenhouse gases and resulting climate change harm the health of people in the United States."
>
> "(4) Changes in climate resulting from human-caused emissions of greenhouse gases harm the welfare of people in the United States."
>
> "(5) Continued emissions of greenhouse gases from human activities will lead to more climate changes in the United States, with the severity of expected change increasing with every ton of greenhouse gases emitted."

Observe that these conclusions are qualitative rather than quantitative. Qualitative characterization of the damages from climate change persists throughout the report. Moreover, the report does not discuss the cost of abatement of emissions. Thus, beyond endorsing the binary EPA Endangerment Finding, the NASEM report provides no basis for evaluation of the costs and benefits of alternative regulations that the EPA might establish.

The qualitative framing of the report applies to its discussion of uncertainty as well.[4] The report uses the word "uncertain" or "uncertainty" multiple times, but almost always in a binary manner. For example, a brief discussion of the feasibility of human adaptation to climate change states (p. 45): "The potential effectiveness of adaptation measures in reducing future climate-driven health risks is uncertain." When discussing uncertainty in climate forecasting, the Committee writes (p. 38): "Climate projections have uncertainty due to internal variability in the climate system, uncertainty in future emissions of GHGs and aerosols, and structural uncertainty in the models themselves."

---

[4] Sustained quantification of uncertainty appears only in Appendix C, which is a 2009 EPA document rather than a product of the 2025 NASEM Committee.

## 3.2. The DOE Report

The report of Climate Working Group (2025) was commissioned by the U.S. Secretary of Energy, the stated objective being (p. viii): “to encourage a more thoughtful and science-based conversation about climate change and energy.” Describing the conclusions of the report, he wrote (p. viii): “many readers may be surprised by its conclusions—which differ in important ways from the mainstream narrative. That’s a sign of how far the public conversation has drifted from the science itself. To correct course, we need open, respectful, and informed debate.”

The main findings of the report are summarized in the Executive Summary. They include (p. ix):

- “Elevated concentrations of $CO_2$ directly enhance plant growth, globally contributing to ‘greening’ the planet and increasing agricultural productivity.”
- “Climate change projections require scenarios of future emissions. There is evidence that scenarios widely used in the impacts literature have overstated observed and likely future emission trends.”
- “The world’s several dozen global climate models offer little guidance on how much the climate responds to elevated $CO_2$, with the average surface warming under a doubling of the $CO_2$ concentration ranging from 1.8°C to 5.7°C . . . . The combination of overly sensitive models and implausible extreme scenarios for future emissions yields exaggerated projections of future warming.”
- “Most extreme weather events in the U.S. do not show long-term trends. Claims of increased frequency or intensity of hurricanes, tornadoes, floods, and droughts are not supported by U.S. historical data.”
- “Both models and experience suggest that $CO_2$-induced warming might be less damaging economically than commonly believed, and excessively aggressive mitigation policies could prove more detrimental than beneficial.”

Comparison of these DOE findings with the NASEM conclusions indicates profound dissonance between the two reports. The dissonance persists when one reads both reports in full.

Whereas the DOE report states that elevated concentrations of CO2 increase agricultural productivity, the NASEM report asserts that (p. 58): "Climate-driven changes in temperature and precipitation extremes and variability are leading to negative impacts on agricultural crops and livestock." Whereas the DOE report optimistically claims that increasing $CO_2$ contributes to the positive phenomenon of "global greening," the NASEM report is negative in its appraisal of the environment impacts.

The DOE report claims that mainstream climate research is overestimating the trends over time in both GHG emissions and in global warming. In contrast, the NASEM report states (p. 13): "Among GHGs, estimates of $CO_2$ emissions have the lowest uncertainties because the majority of these estimates are based on fuel consumption data, which are accurately and precisely tracked, multiplied by the emissions per usage, which is also well-known."

The DOE report finds that "Most extreme weather events in the U.S. do not show long-term trends." In contrast, the NASEM report states (p. 22): "Observations show continuing increases in hot extremes alongside declines in cold extremes" and "In the United States, regional shifts in annual precipitation and a higher number of extreme single-day precipitation events have been observed."

The NASEM report contains no discussion of the cost of mitigation policies. In contrast, the DOE report finds that "excessively aggressive mitigation policies could prove more detrimental than beneficial." Going further, the DOE report seeks to delineate the circumstances in which aggressive policies would be worth the cost. Using the concept of *equilibrium climate sensitivity* (ECS) to measure the magnitude of global warming, the report states (p. 25):

> "Uncertainties in ECS are highly consequential for policy making. . . . economic models use ECS values to project the costs of $CO_2$ emissions. The traditional value (3.0 °C) has typically yielded modest global social costs of $CO_2$ emissions, sufficient to justify some policy actions, but mostly deferred to later in this century. If ECS is very high (above 4.5°C) immediate aggressive emission controls become more imperative, whereas no $CO_2$ emission controls are economically justifiable for ECS below 2.0°C. . . . . Obtaining a precise estimate is impossible, so policy making needs to account for the uncertainty."

The attention given to uncertainty in this passage persists throughout the report. The authors present quantitative measures of uncertainty regarding the ECS and sea level rise drawn from IPCC materials and other mainstream literature. The reports contains a full chapter (Chapter 8) on the topic: "Uncertainties in Climate Change Attribution."

The DOE report ends with "Concluding thoughts" that emphasize the importance of quantitative evaluation of benefits and costs as well as recognition of uncertainties when evaluating policies. The authors state (p. 130):

> "This report supports a more nuanced and evidence-based approach for informing climate policy that explicitly acknowledges uncertainties. The risks and benefits of a climate changing under both natural and human influences must be weighed against the costs, efficacy, and collateral impacts of any 'climate action', considering the nation's need for reliable and affordable energy with minimal local pollution. Beyond continuing precise, uninterrupted observations of the global climate system, it will be important to make realistic assumptions about future emissions, re-evaluate climate models to address biases and uncertainties, and clearly acknowledge the limitations of extreme event attribution studies. An approach that acknowledges both the potential risks and benefits of $CO_2$, rather than relying on flawed models and extreme scenarios, is essential for informed and effective decision-making."

However, this nuanced and neutrally written passage is at odds with the relentless effort made throughout the DOE report to portray mainstream climate research as biased in the direction of promoting overly aggressive climate policies.

It is particularly striking to read the opening statement in the concluding Chapter 12 (p. 130): "U.S. policy actions are expected to have undetectably small direct impacts on the global climate and any effects will emerge only with long delays." The authors go on to call attention to the fact that the 2009 EPA Endangerment Finding has been used as a justification to strengthen regulation of motor vehicle emissions. They argue against such regulation, stating (p. 130): "even the most aggressive regulatory actions on GHG emissions from U.S. vehicles cannot be expected to remediate alleged climate dangers to the U.S. public on any measurable scale."

The DOE report bases this strong anti-regulatory statement on their brief discussion (p. 129) of the "scale problem," which observes that climate change is a planetary rather than a local phenomenon. They interpret the United States as a local entity whose actions are too limited in scope to have a measurable effect on global warming. Mainstream climate research has long recognized the scale problem and has used it to argue for planet-wide action, as exemplified by the 1997 Kyoto Protocol and the 2015 Paris Agreement. Considering the scale problem, the DOE report could have likewise recommended planet-wide action. Instead, it advocated elimination of climate-motivated EPA regulation of GHG emissions.

## 4. Conclusion

And so the climate war continues. Whatever the policy question under consideration, credible analysis is essential. Unfortunately, science and ideology have become increasingly difficult to separate when the public considers climate policy in the United States. Using welfare economics to recognize tradeoffs and face up to uncertainty will not eliminate scientific disagreements and differences of perspective on societal values. Yet It should enable more productive communication across researchers and amongst the public.